\documentstyle[12pt,epsf]{article}
\newcommand{\op}{{\overline\phi}}

\def\half{{1\over2}}
\def\fourth{{1\over4}}
\def\a{{\alpha}}
\def\b{{\beta}}
\def\de{{\delta}}

\def\sech{{\rm sech}}
\newcommand\eref[1]{(\ref{#1})}

\newcommand{\be}{\begin{equation}}
\newcommand{\ee}{\end{equation}}
\newcommand{\ben}{\begin{eqnarray}\displaystyle}
\newcommand{\een}{\end{eqnarray}}
\newcommand{\refb}[1]{(\ref{#1})}

\newcommand{\sectiono}[1]{\section{#1}\setcounter{equation}{0}}

\begin{document}

{}~ \hfill\vbox{\hbox{hep-th/0008231}
\hbox{CTP-MIT-3019}}\break

\vskip 3.5cm

\centerline{\large \bf Field theory models for tachyon and }
\vspace*{1.5ex}

\centerline{\large\bf  gauge field string dynamics}

\vspace*{10.0ex}

\centerline{\large \rm Joseph A. Minahan\footnote{Address
after Sept. 1, 2000:
Department of Theoretical Physics, Uppsala University, Uppsala, Sweden}
 and Barton Zwiebach
}

\vspace*{1.5ex}

\centerline{\large \it Center for Theoretical Physics}
\centerline{\large \it Massachussetts Institute of Technology}
\centerline{\large \it  Cambridge, MA 02139, USA}
\vspace*{2.0ex}
\centerline{\rm e-mail: minahan@mit.edu, zwiebach@mitlns.mit.edu}

\vspace*{4.5ex}
\medskip
\centerline {\bf Abstract}
\noindent 
In hep-th/0008227, the unstable lump solution of $\phi^3$ theory
was shown to have a spectrum governed by the solvable Schroedinger
equation with the $\ell=3$ reflectionless potential
and was used as a model for tachyon condensation in string theory.
In this paper we study in detail an $\ell\to \infty$ scalar field
theory model whose lump solution mimics remarkably
the string theory setup: the original field theory tachyon and the
lump tachyon have the same mass, the spectrum of the lump consists of
equally spaced infinite levels, there is no continuous
spectrum, and nothing survives after tachyon condensation.
We also find exact solutions for lumps with codimension $\ge 2$,
and show that that their tensions satisfy
$\frac{1}{2\pi}\frac{T_p}{T_{p+1}}=  
\frac{e}{\sqrt{2\pi}}\approx 1.08$.
We incorporate gauge fixed  couplings to a $U(1)$ gauge field
which preserve  solvability and
result in 
massless gauge fields on the lump.

\bigskip

\vfill \eject
\baselineskip=17pt

\tableofcontents

\sectiono{Introduction and Summary} \label{s1}

While the precise description of the physics of tachyon condensation
and D-brane annihilation may require the complete framework
of string field theory, it is of interest
to consider simpler models to understand some of the
puzzling issues that arise. Indeed, tachyon condensation
is simpler in the case of $p$-adic open string theory \cite{0003278},
and in the case when large non-commutativity is introduced on the
D-branes
\cite{0005006,0005031,0006071,0007226,0008013,0008064}.

In a recent paper \cite{0008227} a model was developed where, just
as in string field theory, we can explore
the decay of a soliton as seen by the field
theory living on the world-volume of the soliton itself.
The model is based on the unstable lump solution of
the simple $\phi^3$ theory, the
truncation of open string field theory to the tachyon
field only (this lump was first studied numerically in \cite{0002117}).
Not only is the exact lump profile
readily written in terms of hyperbolic
functions, but the full spectrum of the field theory
on the lump world-volume is readily obtained. This happens
because  the potential
in the relevant Schroedinger type equation is the $\ell=3$
case of the infinite series of exactly solvable reflectionless
potentials:\footnote{For a pedagogical   
review on these and other solvable Hamiltonians, with references
to the early literature see \cite{9405029}. Applications of 
reflectionless systems to fermions can be found in \cite{9408120}.} 
$U_{\ell} (x)
= - \ell \,( \ell + 1) \,\hbox{sech}^2 x$.
The potential $U_{\ell}$ gives precisely $\ell$ bound states, 
and a continuum, whose lowest energy state is a bound state at
threshold.
The potentials $U_1$ and $U_2$ were known to be relevant
to the sine-Gordon soliton and the $\phi^4$ kink respectively
\cite{christ}. In both  cases the wavefunction of the
lowest energy bound state in the
quantum mechanical problem is identified with the derivative
of the soliton profile. This procedure does not give a
simple result for $\ell \geq 3$ \cite{christ,GJ}. In the
model of \cite{0008227} it is the middle bound state of the
$U_3$ potential that is identified with the derivative of the lump
profile.
The lowest bound state then represents the tachyonic fluctuation
around the lump. Motivated by this result, one is led naturally to
consider
building lumps for $\ell >3$ by using the next to lowest bound state of
$U_{\ell}$ to define the profile derivative. The scalar field theory
potentials
giving rise to such unstable lumps were found recently \cite{GJaffe}.
The remarkably simple field theory potentials $V_{\ell+1}(\phi)$
(whose lumps are governed by $U_{\ell+1}$) consist
of two terms, a $\phi^2$ term and a $(\phi^2)^{1 + {1\over \ell}}$ term.
For each $V_{\ell+1}$ the lump worldvolume theory will have 
one tachyon, one massless scalar, $(\ell-1)$ massive scalars
and a continuum spectrum.

In this paper we study these field theories as models
for tachyon condensation on unstable branes. Our focus, however,
will be in the model obtained as an $\ell\to \infty$ limit. As
will be explained in the text, by simultaneously letting $\ell\to
\infty$
and rescaling the spacetime coordinates, one obtains the well defined
potential
\be
\label{vinf}
V_\infty (\phi) = -{1\over 4} \phi^2 \ln \phi^2 \,.
\ee
This very unusual potential has an unstable critical point at
$\phi_0^2 =e^{-1}$, with a tachyon mass squared $m^2 =-1$.
Note, in addition,  the local minimum at $\phi=0$,
where $V''\to + \infty$ and the scalar field
acquires infinite mass. With a standard kinetic term for
$\phi$ we have a simple field theory of a tachyon with
the expected property that as the tachyon condenses to the
local minimum it disappears from the spectrum by acquiring
an  infinite mass.

The lump solution in this field theory is a simple gaussian $\op(x)=
e^{-x^2/4}$ and the Schroedinger potential $U_\infty$ 
is that of the one-dimensional simple harmonic oscillator!
The spectrum of the lump field
theory includes a tachyon, a massless field, and an infinite tower
of equally spaced massive scalars. It is noteworthy that the tachyon
living on the lump has $m^2=-1$, just as the mass of the tachyon
in the field theory model giving rise to the lump. This is true
for string theory D-branes 
and for $p$-adic solitonic $q$-branes \cite{0003278} 
but not for the finite $\ell$ models.
Perhaps more significant is the fact that the continuum spectrum
of the finite $\ell$ models has disappeared. Indeed, in ordinary
field theory, the continuous spectrum of the soliton is a reflection
of the degrees of freedom that exist on the vacuum defined by the
asymptotic field expectation value of the soliton at spatial infinity.
For
ordinary potentials, the asymptotic field configuration is a vacuum
with perturbative particle excitations. For the case of \refb{vinf} the
lump is
asymptotic to the $\phi=0$ configuration, around which there is no
perturbative dynamics.

Tachyon condensation on the lump is easily studied. The above
theory has the special feature that the  wavefunction representing
the tachyon mode on the lump
is actually proportional to the lump profile itself.  This means that
out of the infinite number of fields living on the lump, only
the tachyon needs to flow in order to destroy the lump. It can
be seen explicitly how this flow makes all fields on the lump
flow to infinite mass.

\medskip
The massless mode on the lump is just a derivative of the lump
profile.  Hence, if this mode is given a small expectation value, 
then
the lump will have a correspondingly small shift in position.  This mode
behaves as a marginal operator in string field theory.
Just as in the $\ell=3$ model, we confirm that in the $\ell=\infty$
model
the relation between displacement and the marginal 
parameter is two to one.
Both small displacements and large displacements lead to a small
marginal
parameter, with the parameter taking a finite maximum value in
between. In the present model, however, we are able to see explicitly
how
higher level states help realize large displacements;
for a large shift $x_0$, the largest
contribution to the shift comes from modes with level number $L$
satisfying
$L\sim x_0^2/4$. These results provide evidence for the missing
branch proposal of \cite{0008227} which suggests that a different
choice of branch for higher level fields describes large marginal
deformations in string field theory. 
The branch point occurs at the maximal value
for the string field marginal parameter \cite{0007153}.

\medskip
There is another string theory parallel. For the $\ell=\infty$
model we can find the profiles of all lower codimension lump solutions
explicitly. Thought of as branes of various spatial dimensions, we
find
that the tensions satisfy ${1\over 2\pi} {T_p\over T_{p+1}} = {e\over
\sqrt{2\pi}}$($\simeq 1.084)$, for all possible $p$. In string theory
this ratio is unity. A similar property was found to hold in
$p$-adic string theory, where the above ratio depends on the
prime number used to define the model.

\medskip
We also extend  the models to incorporate
a $U(1)$ gauge field. The terms we add were suggested
by the form of level expanded string theory, but our
main constraints come from requiring: (i) solvability of the
spectrum, (ii) inheriting an exactly massless gauge field
on the worlvolume of the lump and (iii) existance of the
model for all finite $\ell$. These conditions can all
be satisfied and  lead to
some special forms for scalar-gauge field interactions.
It is possible, in addition, to guarantee that the gauge field
components
transverse to the brane do not give rise to additional
massless states. This is required because the translation mode of the
brane
arises from the tachyon field. The lagrangians we write should be
thought as gauge fixed ones, just as level expanded versions
of string theory are generally worked out in the Siegel gauge.
We do not know if there is a gauge invariant version of the models.

\medskip  
This paper is organized as follows. In section \ref{s2} we introduce
the field theory scalar models associated with
arbitrary $\ell$ and
infinite $\ell$. We also calculate the spectrum of the lump worldvolume
field theories in both cases. In section \ref{s3} we show explicitly
how in the $\ell=\infty$ model tachyon condensation sends the masses
of all fields living on the lump to infinity. In section \ref{s4} 
we discuss
large displacements of $\ell=\infty$ lumps, and study an effective
field theory approximation where we include only the tachyon and
the massless field on the lump. In section 5 we study higher 
codimension
lumps, give their exact profiles and calculate their tensions.
In section 6 we show how to incorporate gauge fields into the
models. Some concluding remarks are offered in section 7.

\sectiono{Finite $\ell$ models and the $\ell\to \infty$ model}\label{s2}

In this section we construct a series of field theory models whose lump
profiles are exactly solvable.  Moreover, the spectrum of fluctuations
about the lumps is also solvable.  In the first subsection, we review
a construction of Goldstone and Jaffe \cite{GJaffe} for building field
theory potentials giving rise to 
lumps with spectra controlled by 
the $U_{\ell+1}$ reflectionless potentials. We then construct a new
potential by
taking $\ell\to\infty$, and investigate various properties of this
potential. In
the second subsection, we discuss the spectrum of lump fluctuations.

\subsection{Introducing the models}
As our starting point, let us consider the one dimensional Schroedinger
equation for the $U_{\ell+1}$ potential,
\begin{equation}\label{rpseq}
-\frac{d^2 \psi(x)}{d x^2}  
-(\ell+1)(\ell+2)\,\sech^2(x)\psi(x) = E\,\psi(x).
\end{equation}
If $\ell$ is an integer, then the potential is reflectionless, in other
words
a wave traveling in from the right will be completely transmitted when
crossing over the potential.
This Schroedinger equation in \eref{rpseq} is completely solvable.  The
ground
state wave function has the solution
\begin{equation}\label{gswv}
\psi_0(x) = \frac{1}{\cosh^{\ell+1}(x)}\, ,
\end{equation}
with energy
\begin{equation}\label{gsE}
E_0 = -(\ell+1)^2 \,.
\end{equation}
There are a finite number of bound solutions, whose wavefunctions are
the
ground state wavefunction in \eref{gswv} multiplied by Hermite
polynomials
in $\sinh(x)$.  The energy levels for these bound states are
\begin{equation}\label{bsel}
E_n=-(\ell+1-n)^2\,,\qquad 0\le n < \ell+1.
\end{equation}
Notice that $n=\ell+1$ would give zero energy. Indeed at the
bottom of the continuum there is a bound state at threshold. This state
will not play an important role in all that follows.

\medskip 
The field theory potential giving rise to a lump solution
with fluctuation spectrum governed by the Schroedinger equation
\eref{rpseq} was found in \cite{GJaffe}\footnote{We thank
J. Goldstone and R. Jaffe for informing us of this result
prior to publication.}.
It is given by
\begin{equation}\label{poteq}
V_{\ell+1}(\phi)=\frac{\ell}{4}\phi^2(1-\phi^{2/\ell}).
\end{equation}
For $\ell >2$  
we define
$\phi^{2/\ell}\equiv (\phi^2)^{1/\ell}$ with the $\ell^{\rm th}$ root
real and positive.

Let us review the construction of the  potential in \eref{poteq}.
We assume the following Lagrangian for the
field theory:
\begin{equation}\label{lag}
{\cal L}_{\ell+1} =  
-\,\half \partial_\mu\phi\partial^\mu\phi - V_{\ell+1}(\phi)\, , 
\end{equation}
where $V_{\ell+1}(\phi)$ is to be determined, and we work
with the metric $(-, + \cdots +)$.   
A lump solution to the
equations of
motion satisfies
\begin{equation}\label{lump}
{\op\,}''(x)={V_{\ell+1}}'(\op(x)),
\end{equation}
which, after integrating is
\begin{equation}\label{com}
\half({\op\,}'(x))^2=V_{\ell+1}(\op(x)).
\end{equation}
An integration constant has been absorbed into the definition of
$V_{\ell+1}(\phi)$, making the potential vanish at the asymptotic
value taken by the field in the lump solution. 
Now assume that the {\it derivative} of $\op(x)$
is proportional to the first
excited state wavefunction of \eref{rpseq},
\begin{equation}\label{zeromode}
\op'(x)=-\sqrt{\frac{\ell}{2}}\,
\frac{\sinh(x/\sqrt{2\ell})}{\cosh^{\ell+1}(x/\sqrt{2\ell})},
\end{equation}
where the argument of the cosh and sinh functions has been rescaled for
later convenience.
  The derivative acting on $\op$ generates a translation
of the lump, hence the wavefunction in \eref{zeromode} is a
zero mode of the fluctuation spectrum.  The expression in
\eref{zeromode}
is easily integrated, giving
\begin{equation}\label{lumpsol}
\op(x) = \frac{1}{\cosh^\ell(x/\sqrt{2\ell})}\, .
\end{equation}
We can now use standard  identities to express
$(\op'(x))^2$
in terms
of $\op$,
\begin{equation}\label{phirel}
({\op\,}')^2=\frac{\ell}{2}\,\op^{\,2}\,(1-\op^{\, 2/\ell}).
\end{equation}
Hence, using \eref{com} the potential is given by \eref{poteq}.

The potential is not bounded from below, thus we expect the fluctuations
about the lumps to have tachyonic modes. 
$V_2(\phi)$  is an inverted quartic potential.  $V_3(\phi)$
is the $\phi^3$ potential that
describes the level zero string field
theory and the model studied in detail in \cite{0008227}.  
For $\ell>2$, the
potential $V_{\ell+1}(\phi)$ is nonpolynomial.

The potential $V_{\ell+1}(\phi)$ has a maximum at
\begin{equation}
\phi=\phi_0=\left(\frac{\ell}{\ell+1}\right)^{\ell/2} \,,
\end{equation}
and a local minimum at $\phi=0$.  At the maximum, the second derivative
of the potential is given by
\begin{equation}\label{Vmax}
V''(\phi_0)=-1\, ,
\end{equation}
while at the minimum, it is
\begin{equation}\label{Vmin}
V''(0)=\frac{\ell}{2}\, .
\end{equation}
Given the general behavior of these potentials, it is tempting to use
one of the higher $\ell$ models as a toy model of open string field
theory.
The analog of the open string vacuum is  the maximum $\phi=\phi_0$, 
with an
open string tachyon of mass squared $m^2=-1$.
The closed string vacuum analog is the local minimum at $\phi=0$. 

\begin{figure}[!ht]
\leavevmode
\begin{center}
\epsfbox{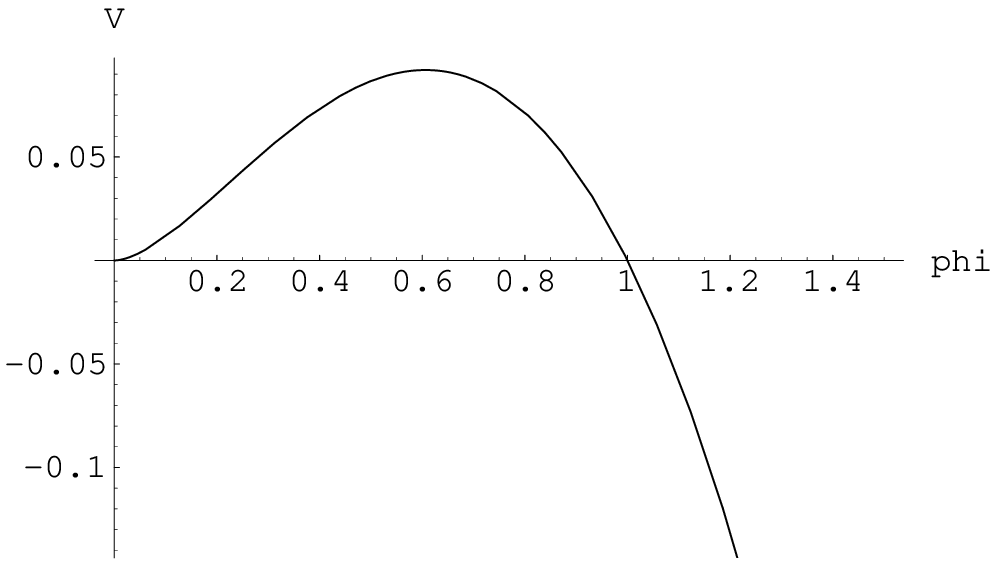}
\end{center}
\caption[]{\small The $\ell\to \infty$ potential $V_\infty(\phi) =
-{1\over 4} \phi^2 \ln \phi^2$.}
\label{f1}
\end{figure}

\begin{figure}[!ht]
\leavevmode
\begin{center}
\epsfbox{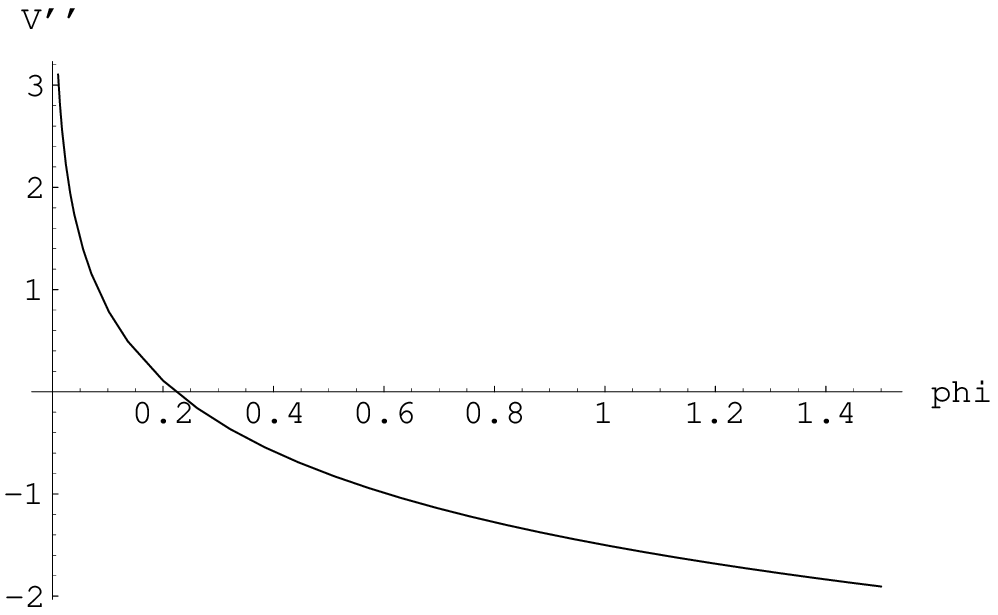}
\end{center}
\caption[]{\small The second derivative ${{V_\infty}}''(\phi) =
-{1\over 2} (3 + \ln \phi^2)$.} \label{f2}
\end{figure}

At the local minimum,  the scalar field has  a mass squared
$m^2=\ell/2$, hence in
the limit $\ell\to\infty$, the scalar field decouples.  However, the
tachyon
mass at the maximum remains finite.  Hence, the $\ell\to\infty$  model
 captures some of the salient features of open string field theory.
In this limit,
the potential in \eref{poteq} simplifies to
\begin{equation}\label{Vinf}
V_\infty(\phi)=-\fourth\phi^2\ln\phi^2,
\end{equation}
which
 has a maximum at $\phi_0=e^{-1/2}$. This potential is shown in
figure \ref{f1}.  Its higher derivatives are also of interest.  In
particular,
note that
\begin{equation}\label{Vinfdd}
{V_\infty}''=-\half\ln\phi^2-{3\over 2}\,,
\end{equation}
which blows up as $\phi\to 0$.  Hence the effective scalar 
mass blows up
as $\phi$ flows to the analog of the closed string vacuum.
This is shown in figure \ref{f2}.

We can easily find the lump solution for the $\ell\to\infty$ case by
solving
\eref{com} by quadratures.  This gives
\be
\label{quad}
x=\int_\op^1 \frac{d\phi}{\sqrt{2V_\infty(\phi)}}
=2\,\sqrt{-\ln\op} \quad \to \quad \op(x)=\exp(-x^2/4),
\ee
a simple gaussian.  This is reminiscent of the lump solutions for the
$p$-adic string field theory \cite{0003278}.

\subsection{Field theory spectrum on the lumps} 

We now consider the spectrum for the codimension one lumps of these
models, both for finite $\ell$ and in the $\ell\to\infty$ limit.
If we expand $\phi$ about the lump solution $\op$, then the fluctuation
modes
satisfy the Schroedinger equation
\begin{equation}\label{fmodes}
-\,\frac{d^2\,\psi(x)}{dx^2}+{V_{\ell+1}}''(\op(x))\psi(x)=m^2\psi(x).
\end{equation}
Substituting \eref{poteq} and \eref{lumpsol}
into \eref{fmodes}, the equation becomes
\begin{equation}\label{fmodes2}
-\frac{d^2}{dx^2}\psi(x)+\frac{1}{2\ell}\left(\ell^2-
(\ell+1)(\ell+2)\sech^2(x/\sqrt{2\ell})\right)\psi(x)=m^2\psi(x).
\end{equation}
Except for the shift and rescaling of the potential, and the
corresponding
rescaling of $x$, this is the reflectionless potential in \eref{rpseq}.
We can then find the masses by shifting and rescaling \eref{bsel},
leaving us
with
\begin{eqnarray}\label{masses}
m^2&=&\frac{1}{2\ell}\left(\ell^2-(\ell+1-n)^2\right)\,,\qquad
0\le n < \ell+1   
\nonumber\\
   &=& \frac{(n-1)(2\ell-n+1)}{2\ell}\, .
\end{eqnarray}

{}From \eref{masses} we see that the lowest mode has
negative $m^2= -1 -{1\over 2\ell}$ 
and that the next mode is massless.  We also see that there
are $\ell-1$ discrete massive modes
and a continuous spectrum starting at $m^2=\ell/2$.

If we take the $\ell\to\infty$ limit, then \eref{masses} reduces to
$m^2=n-1$ and the continuum is pushed to infinite mass.  
We can see this
explicitly by inserting the potential $V_\infty(\phi)$ in \eref{Vinf}
and
the lump solution in \eref{quad}
into \eref{fmodes}.
Then the Schroedinger equation for the fluctuation modes
reduces to the harmonic oscillator equation
\begin{equation}\label{fmodesinf}
-\frac{d^2}{dx^2}\psi(x)+\left(-\frac{3}{2}+
\fourth x^2\right)\psi(x)=m^2\psi(x).
\end{equation}
The masses of the modes are well known and  satisfy
\be
\label{masss}
m^2=n-1\,,  \quad n\ge0 \,.
\ee
Hence, unlike the finite $\ell$ models, the $\ell\to\infty$ model
shares three  important properties with string
theory.  First, the fluctuations about the lumps have only
a discrete spectrum. Second, the mass of the tachyon fluctuation equals
the mass of the tachyon in the original field theory. Third, there is
equal spacing between levels. Since in string theory we assign
level zero to the tachyon we do the same here and define the
level $L(n)$ to be 
\be
\label{leveldef}
L (n) = n\,.
\ee

The lump solutions have a profile in which the field
falls off to the closed
string vacuum as $|x|\to\infty$.
Examining \eref{fmodes}, we see that a
necessary
condition for a discrete spectrum is that $V''\to\infty$ as
$|x|\to\infty$.  But this is simply the statement that the
scalar mass becomes
infinite as the scalar field descends towards
the analog of 
the closed string vacuum.
In fact, the discrete spectrum condition should be true
for models that include higher modes in the level expansion, 
and even the complete open string field
theory.  Hence, the decoupling of the open string tachyon as a
propagating
mode is consistent
with the fact that the open string spectrum of D branes is discrete.

\sectiono{Tachyon condensation 
and mass flow at  $\ell= \infty$} \label{s3} 

In this section we explore how the lumps decay through tachyon
condensation.
In the limit that $\ell\to\infty$, the process is especially simple.

In the
previous section, we  learned that the finite $\ell$ models have lump
profiles of the form $\op(x)=\sech^{\ell}(x/\sqrt{2\ell})$.  However,
the tachyon fluctuation of the lump has a wavefunction
$\psi_0(x)=\sech^{\ell+1}(x/\sqrt{2\ell})$.  When the lump decays, the
fluctuations
become large, such that the lump profile is exactly cancelled by the
fluctuations.  However, since the profile wavefunction is not quite the
same
as the tachyon fluctuation, other modes must be turned on in order to
annihilate the lump.  This was explored in great detail for the
$\ell=2$ case in \cite{0008227}.

However, when $\ell$ approaches infinity, the tachyon wavefunction
is the lump profile.  We can see this explicitly from the
$\ell\to\infty$
Schroedinger equation in \eref{fmodesinf}.  Since this is the equation
for
a harmonic oscillator, the ground state wavefunction
{\it is} the gaussian $\psi_0(x)=e^{-x^2/4}$, which is the lump profile
equation.  Hence, when the lump decays, only the tachyon  mode
condenses.

Let us show more explicitly that the tachyon flow does not
induce condensation for the other modes.
To this end, write the scalar field $\phi$ as 
\begin{equation}\label{tachyon}
\phi =(1+T)\, \op(x) + \psi(x) =(1+T) \, e^{-x^2/4} + \psi(x) \,,
\end{equation}
where the first term is an expectation value, and $\psi(x)$ denotes
fluctuations about it. The lump
has decayed when $T$ reaches $T=-1$. If the other fields 
do not condense under the flow, then
their one point functions in the effective Lagrangian must be seen
to vanish. 
Expanding the action corresponding to \refb{lag}  with $\ell=\infty$
about the decaying lump described in \refb{tachyon}
and using the equations of
motion, the one point contributions are
\begin{eqnarray}\label{1ptfn}
&&\int dx [-(1+T){\partial_x}^2 \op(x)+{V_\infty}'((1+T)\op(x))]\psi(x)
\nonumber\\
&&\qquad\qquad=\int dx [-(1+T){V_\infty}'(\op(x))+
{V_\infty}'((1+T)\op(x))]\psi(x),
\end{eqnarray}
where $\psi(x)$  includes all fluctuation modes. 
Using the explicit
form for $V_\infty(\phi)$ in \eref{Vinf}, we find that the term inside
the square brackets is $-\op(x)\ln(1+T)$.  Since $\op(x)$ is
proportional to
the ground state wave function and except for its tachyon
component $\psi(x)$ is orthogonal
to the ground state,
the integral over $x$ is zero.  Hence the one point functions of all
modes other than the tachyon are zero.

Next, let us examine what happens to the masses as the tachyon
condenses.
The effective mass for the modes is found by solving the eigenvalue
equation
\begin{equation}\label{effmass}
-\frac{d^2\,\psi(x)}{dx^2}+{V_\infty}''((1+T)\op(x))\,\,\psi(x)=
m^2\psi(x).
\end{equation}
Using the relation
\begin{equation}
{V_\infty}''((1+T)\op(x))={V_\infty}''(\op)-\ln(1+T)\,,
\end{equation}
we see that \eref{effmass} is the same as \eref{fmodesinf}  with a
finite shift in the mass, $\Delta m^2=-\ln(1+T)$.  As $T\to-1$, this
shift
blows up and the modes become infinitely massive, 
effectively decoupling from
the spectrum.

\sectiono{Transporting the  $\ell=\infty$ lumps}\label{s4}

Because of translational invariance, the lumps have zero modes.
If the zero mode is given an infinitesimal expectation
value, the position of the lump changes infinitesimally. Such
a zero mode is quite analogous to the modes associated to
marginal operators in string theory.
The lump zero mode alone, however, cannot implement large
translations-- this is the statement that large marginal
deformations require giving expectation values to other fields.
This was explored in some detail for string field theory in
\cite{0007153} and for the   $\phi^3$ toy model in \cite{0008227}.

In particular the focus in
\cite{0007153} and \cite{0008227} was on a special phenomenon.
In string field theory, the effective potential for
the marginal parameter is only defined up to a critical
value of this parameter \cite{0007153}. An explanation
for this was proposed in \cite{0008227}. The idea is that
as the displacement of the lump increases from zero to
infinity, the marginal
parameter first grows, reaches a maximum and then decreases down to
zero. Higher level fields take over in displacing the lump as the
displacement grows.

Evidence for this proposal was given in the $\phi^3$ model, but
there the role of the higher level fields was taken by the
continuum of scalar fields. In the present $\ell \to \infty$ model
there is no continuum, and we can see explicitly how the higher
level fields, all corresponding to localized bound states do the job of
moving the lump. Thus, the $\ell\to\infty$ model is expected
to model more closely the behavior of marginal parameters in 
string field theory.

We can see this for the $\ell\to\infty$ model
 by decomposing the
shifted lump into the fluctuation modes of the unshifted lump.
The shifted lump  has a profile
\begin{eqnarray}\label{shiftlump}
\op(x-x_0)&=&\exp(-(x-x_0)^2/4)\nonumber\\
          &=&\sum_{n=0}^\infty A_n \psi_n(x),
\end{eqnarray}
where $\psi_n(x)$ are the harmonic oscillator wavefunctions.
In this notation the $A_n$'s are the fields living on the
lump. The wavefunctions, given by
\be
\label{showf}
\psi_n (x) = {1\over 2^{n/2} \sqrt{n!} } H_n \bigl( {x\over \sqrt{2}}
\bigr)  e^{-x^2/4}\,,
\ee
are normalized to
\begin{equation}
\int dx \psi_n(x)\psi_m(x)=\sqrt{2\pi}\, \de_{mn}.
\end{equation}
Since $\psi_0(x) = \op (x)$, the tachyon field $T$ representing
a fluctuation of the lump is related to $A_0$ as $A_0 = 1+T$. The
coefficients
$A_n$ can be found in a standard quantum mechanics text
\cite{Schiff} and are given by
\begin{equation}\label{Anrel}
A_n (x_0)=\frac{{x_0}^n}{2^n \sqrt{n!}} \exp(-{x_0}^2/8).
\end{equation}

\begin{figure}[!ht]
\leavevmode
\begin{center}
\epsfbox{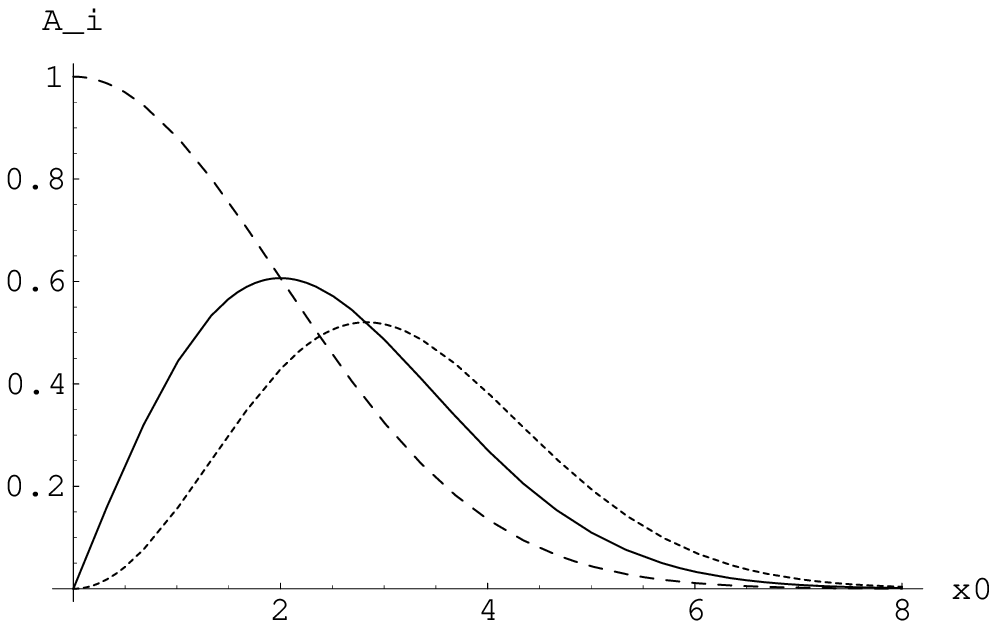}
\end{center}
\caption[]{\small The dashed line is the behavior
of $A_0 = 1+T$ as a function of the lump displacement
$x_0$. The continuous line shows the marginal mode
$A_1(x_0)$, which takes a maximum value for $x_0=2$.
The dotted line is $A_2(x_0)$, which takes a maximum
value after $A_1$ does.} \label{f3}
\end{figure}

\medskip
Clearly, shifting the lump turns on more than just the massless mode
$A_1$, it turns on all fields.   For small $x_0$, we have $A_1 \sim
x_0/2$, but
$A_1$ takes a maximum value for $x_0=2$:
\be
\label{maxa1}
A_1^{max} = A_1 (x_0=2) = \exp(-1/2) \simeq 0.6065\,.
\ee
$A_1(x_0)$ decreases for $x_0>2$ beyond.  A plot showing
the behavior of $A_0, A_1$ and $A_2$ as functions of $x_0$
is shown in Fig.~\ref{f3}. For $x_0=2$, the lump constructed
using  the  four
lowest modes $(A_0$ up to $A_3$) is centered at about
$x_0=1.9$. Using only $A_0$ and $A_1$ the lump ends up centered
at $x=1$.

Examining the $n$-dependence of the coefficients
$A_n$ for large $x_0$ one can show that they become largest for $n\sim
x_0^2/4$. Since the level $L$ of a field is simply given
as $L(n) =n$ (see \refb{leveldef}), we see that
that in displacing the lump a large distance
$x_0$, fields of level
\be
L = {x_0^2\over 4}\,,
\ee
provide the largest contribution.

We have attempted to reproduce the above qualitative
behavior by constructing the effective potential of
the tachyon and marginal mode only.  Let $\a$
denote the zero mode amplitude and $T$ denote
the tachyon amplitude. In other words, we set
\be
\label{fcon}
\phi (x) = (1+T + \alpha \, x) \exp (-x^2/4) \,.
\ee
Note that $\alpha = A_1$ since $\psi_1 (x) = x\exp (-x^2/4)$.
The effective potential for these modes is
\vfill
\eject
\begin{eqnarray}\label{effpot}
V_{eff}(\alpha, T)&=&\int dx \Biggl\{(1+ T)^2\,\frac{x^2}{8} +
\frac{\a^2}{8}\Bigl(1-{x^2\over 2}\Bigr)^2\nonumber\\
&&\qquad
-\fourth(1+T+\a x)^2
\left(\ln(1+T+\a x)^2-\half x^2\right)\Biggr\}\exp(-x^2/2).
\end{eqnarray}
For a given $\a$ we want to find the $T$ that minimizes $V_{eff}$.
In all previous models, one finds that the equation of motion
of $T$ cannot be satisfied for values of $\a$ greater than a
critical value. Taking
the $T$ derivative of \eref{effpot} gives
\begin{equation}\label{deffpot}
\frac{dV_{eff}}{dT}=-\half\int dx (1+T+\a x)\ln(1+T+\a x)^2\exp(-x^2/2).
\end{equation}
The expression in \eref{deffpot} is clearly zero if $T=-1$.
This corresponds to the expectation value that erases the lump.
There can be another zero for a value of $T \in [-1, 0]$.
In fact, for small $\alpha$ one can convince oneself that
\eref{deffpot} has a zero for $T$ slightly below zero. In general
this second solution exists as long as ${V_{eff}}''|_{T=-1}<0$.

Hence, the critical value $\a$ is defined by the condition that
${V_{eff}}''|_{T=-1}=0$.  Taking another derivative on \eref{deffpot},
we find
\begin{eqnarray}\label{ddeffpot}
\frac{d^2V_{eff}}{dT^2}\bigg|_{T=-1}&=&
-\int dx \left[1+\half\ln(\a x)^2\right]\exp(-x^2/2)\nonumber\\
&=&-\sqrt{2\pi}\left(1-\frac{\gamma+\ln2}{2}+\ln\a\right),
\end{eqnarray}
where $\gamma$ is the Euler constant.  Hence, the maximum value for
$\a$ is reached when
\begin{equation}\label{amax}
\a=\a_{max}=\sqrt{2}\exp(-1+\gamma/2)\approx 0.6943\,,
\end{equation}
in decent agreement with \refb{maxa1}. We note, however,
that for $\a_{max}$ we do not get in this model a reasonable
picture of a displaced lump. Indeed, the marginal branch
starts at $T=0$ for $\a=0$ and flows to $T=-1$ at
$\a=\a_{max}$. For values of $\a \leq 0.6$ the resulting
field configuration resembles that of a displaced lump, but not
at the endpoint of the branch, as can be seen in \refb{fcon}. This, we
believe is an intrinsic limitation of this two-field model.

There is, in addition, the branch where $T=-1$, for any
$\a$. It corresponds to fluctuations along the zero mode
of the lump around the stable vacuum. The marginal
branch merges into this branch for the critical value of $\a$.

\sectiono{Higher codimension lumps and descent relations in
the $\ell=\infty$ model}\label{s5} 

The energy density of the codimension one
brane defined by the lump solution (a $p$-brane) is
 given by
\be
E = \int d^{p}\hskip-1pt \vec y\, dx
\Bigl[  {1\over 2} \Bigl(
{d\op\over dx}\Bigr)^2 + V(\op)\Bigl]  = (\hbox{Vol}_y) \int dx \Bigl(
{d\op\over dx}\Bigr)^2 \,.
\ee
With the lump profile $\op (x) = \exp( -x^2/4)$ one readily finds
\be
\label{energylump}
T_p = {E\over (\hbox{Vol}_y)} =\,  {1\over 4} \sqrt{2\pi} \,,
\ee
for the tension of the $p$-brane. Given that the original
unstable vacuum $\phi_0^2 = e^{-1}$
is supposed to represent the
space-filling ($p+1$)-brane, we have $T_{p+1} = V(\phi_0) = 1/(4e)$.
Therefore
\be
\label{ratio}
{1\over 2\pi}\,\, {T_p\over T_{p+1}} = {e\over \sqrt{2\pi}} \simeq
1.084\,,
\ee
a ratio that in string theory takes the value of unity.

We now consider higher codimension lumps and show that (i) their
profiles can be found exactly, and (ii) the ratio in \refb{ratio}
holds for all values of $p$. Let $c$ denote the value of the
codimension ($c=1$ for the one dimensional lump $\op (x) = \exp(
-x^2/4)$).
The differential equation for the profile takes the form
\be
\label{deprof}
{d^2 \phi\over d\rho^2} + \Bigl( {c-1\over \rho}\Bigr)
{d \phi\over d\rho} - V'(\phi) = 0,
\ee
where $\rho = (x_1^2 + \cdots x_c^2)^{1/2}$ is the radial coordinate
transverse to the brane. The relevant solution of this differential
equation
is easily found:
\be
\label{clump}
\op(\rho) = \exp \Bigl( - {\rho^2\over 4} + {1\over 2} (c-1) \Bigr)
\ee
Indeed the value $\op (0)$ at the core of the lump increases
from one, the value for $c=1$, as the codimension is
increased.  This is in accord with the intuition that the second
term in \refb{deprof}, in the mechanical analogy of motion in
a potential $-V$,  represents a
friction term that is overcome by letting the field at the
core have a larger expectation value. Higher codimension lumps
exist, despite Derrick's no-go theorem, because the potential
$V$ is not bounded below.

Having the expression for the lump we can calculate its tension.
Letting $E_c$ denote the energy of the codimension $c$ lump, we have
\be
E(c )  = (\hbox{Vol}_y) \int d^cx \Bigl[
 {1\over 2} \Bigl(
{d\op\over d\rho}\Bigr)^2 + V(\op)\Bigl] \,.
\ee
This gives, for the tension $T(c)$ of the codimension $c$ brane
\be
T(c) \equiv {E(c ) \over \hbox{Vol}_y} ={1\over 4}\, e^{c-1}  \int d^cx
[
 \rho^2-(c-1)]e^{-\rho^2/2} \,.
\ee
Using $\int d^c x = \hbox{Vol} (S_{c-1}) \int_0^\infty d\rho
\rho^{c-1}$,
with $\hbox{Vol} (S_{c-1})={2\pi^{c/2}\over \Gamma(c/2)}$, the integral
is readily done and gives:
\be
T(c)  ={1\over 4}\, e^{c-1} (2\pi)^{c/2}.
\ee
This result, for $c=0$ agrees with the value $T_{p+1} = 1/(4e)$,
and for $c=1$ agrees with the value of $T_p$ in \refb{energylump}.
Finally, we see that it implies that the ratio in \refb{ratio}
is the same for any pair of branes whose codimension differs by
one unit.  One can show that the spectrum of fields living
on  the  codimension $c$ lump is governed by the Schroedinger
potential of the $c$-dimensional simple harmonic oscillator.

\sectiono{Adding gauge fields}\label{s6}

In this section we consider a possible scenario for gauge fields in our
model.
For components along the brane, there seems to be a neat generalization
of
the gauge fixed action \cite{KS} for all values of
$\ell$.
For
components perpendicular to the brane, the generalization does not
appear to
be as nice, although it improves in the limit $\ell\to\infty$.

The gauge fixed string field action is consistent because there
is a BRST symmetry. Our philosophy is to include terms in the action
that are similar to terms found in gauge fixed string field theory and
save
questions about BRST invariance for future work. Another possibility
would be to try to obtain a gauge invariant formulation of the models.
Experience in string field theory, however, suggests that this may
require the inclusion of auxiliary fields \cite{WITTENBSFT}.

\subsection{The dynamics of the gauge field on the brane}

In string field theory, the gauge-fixed action has the
cubic term of the form 
\begin{equation}\label{gaugemass}
\phi A_\mu A^\mu.
\end{equation}
In level zero string field theory,  $\phi$ is proportional to
$V''(\phi)-V''(0)$, since $V(\phi)$ is cubic.  Hence, a possible
generalization
to the mass term in \eref{gaugemass} is
\begin{equation}\label{newgm}
\beta\, (V''(\phi)-V''(\phi_0))A_\mu A^\mu,
\end{equation}
where $\phi_0$, representing the open string vacuum,  is at the maximum
of the
potential and
$\beta$ is a constant.
At the very
least, the gauge fields should be massless when $\phi=\phi_0$.
For the solvable $\ell$ models, we choose 
$\b=-\half\frac{\ell}{\ell+2}$,
in which case the term in \eref{newgm} can be reexpressed as
\begin{equation}\label{newgm2}
{\cal L}_{\ell+1} = -\,\half\partial_\mu A_\nu\partial^\mu A^\nu
-\half
\frac{{V_{\ell+1}}'(\phi)}{\phi}A_\mu A^\mu + \cdots \,, 
\end{equation}
where we have included the gauge fixed kinetic term for the
vector field. Note that for the $\phi^3$
($V_3$) model, the coefficient
$\beta$ differs from
the corresponding coefficient in the string field theory\cite{KS}. 

Now consider $A_\mu$ fluctuations, where $\mu$ is an index 
along the brane.
To this end, we write
$A_\mu={\cal A}_\mu(y)\eta(x)$
where $y$ refers to coordinates on the brane worldvolume. 
Given the interaction term
in \eref{newgm}, the masses of the gauge fluctuations are found by
solving the
eigenvalue equation
\begin{equation}\label{gaugefl}
-\frac{d^2\,
\eta(x)}{dx^2}
+\frac{{V_{\ell+1}}'(\op(x))}{\op(x)}\,\eta(x)=m^2\,\eta(x).
\end{equation}
Using the explicit solutions in \eref{lumpsol} and \eref{poteq},
\eref{gaugefl}
becomes
\begin{equation}\label{gaugefl2}
-\frac{d^2\,\eta(x)}{dx^2}+\frac{1}{2\ell}[\ell^2-\ell(\ell+1)
\sech^2(x/\sqrt{2\ell})]\,\eta(x)=m^2\,\eta(x).
\end{equation}
Hence, just as in the scalar field case, the fluctuations satisfy
the Schroedinger equation for a reflectionless potential.
However, in the gauge case, this potential is $U_\ell$ and not
$U_{\ell+1}$.
Hence, using \eref{bsel} and rescaling,
 the possible discrete values for the masses are
\begin{eqnarray}\label{massgauge}
m^2&=&\frac{1}{2\ell}[\ell^2-(\ell-p)^2]\qquad\qquad 0\le p <\ell
\nonumber\\
&=&\frac{(n-1)(2\ell -n +1)}{2\ell}\qquad\qquad 1\le n <\ell+1. 
\end{eqnarray}
Thus,  except for the absent tachyon,
 the gauge fields have a spectrum that is the same as
the scalar fields.
Without the tachyon, the lowest level fluctuation is  massless.

In the $\ell\to\infty$ limit, the term in \eref{newgm2} reduces to
\begin{equation}
\fourth \left(\ln\phi^2+1\right)A_\mu A^\mu.  
\end{equation}
The mass squared levels are equally spaced and match up with the scalar
field mass levels.  Again, the lowest state is massless.  Clearly, when
$\phi$ rolls to the closed string vacuum, the mass of the gauge field
diverges.

\subsection{Giving a mass to the gauge field transverse to the brane}

One of the more mysterious questions in open string field theory is what
happens to  gauge field components that are polarized transverse to the
brane.
There cannot be zero modes associated with them, since as we
have seen in the previous sections, the zero mode is accounted for by
a zero mode in the scalar field.

In the gauge fixed action
there are, in addition
to the term discussed in the previous subsection, terms of
the form \cite{KS}
\begin{equation}\label{ddphiAA}
\left(\partial_\mu\partial_\nu\phi\right) A_\mu A_\nu.
\end{equation}
In our toy models, we will assume that there is a similar term, except
that
$\phi$ is replaced with $-\ln\phi$.  Note for the $\ell\to\infty$ model,
$-\ln\phi$ is quite natural, since
${V_\infty}'(\phi)/\phi=-\ln\phi-1/2$.
For the finite $\ell$ models, the $\ln\phi$ term seems more {\it ad
hoc}.
Nor can we justify this term by gauge invariance.  We simply
assume that this belongs to the gauge fixed action.\footnote{The string
field
action has other terms quadratic in the gauge fields, although these
extra terms
all have
$\partial_\mu A^\mu$ factors.  One can not really  get rid of these
terms by going
to a Lorentz gauge, since in principle
this action is already gauge fixed.  These extra
terms seem to ruin the solvability of the gauge fluctuations.
Without any further justification, we do not include these types of
terms.}
Thus, the final
Lagrangian quadratic in the gauge fields is
\begin{equation}\label{finallag}
{\cal L}_{\ell+1} = - \half\partial_\mu A_\nu\partial^\mu A^\nu -
\half \frac{{V_{\ell+1}}'(\phi)}{\phi}A_\mu A^\mu 
+\left(\partial_\mu\partial_\nu\ln\phi\right) A^\mu A^\nu. 
\end{equation}

The term in \eref{ddphiAA} {\it does} change the mass spectrum for the
modes
of the transversely polarized gauge field.  
Using the lump solution 
\refb{lumpsol} we find    
\begin{equation}\label{ddlump}
\frac{d^2}{dx^2}\ln\op(x)=-\half\sech^2(x/\sqrt{2\ell}).
\end{equation}
Hence, on account of \refb{finallag} we find that for the transverse
components equation \refb{gaugefl2} 
is modified to
\begin{equation}\label{transflucts}
-\frac{d^2\eta (x)}{dx^2}  
+\frac{1}{2\ell}[\ell^2-\ell(\ell-1)
\sech^2(x/\sqrt{2\ell})]\eta(x)=m^2\eta(x).
\end{equation}
So again we find the Schroedinger equation for a reflectionless
potential,
but now this is the $U_{\ell-1}$ potential.  
Hence the masses of the bound modes are
\begin{eqnarray}\label{masstrans}
m^2&=&\frac{1}{2\ell}[\ell^2-(\ell-1-p)^2]\qquad\qquad 0\le p<\ell-1
\nonumber\\
&=&\frac{(n-1)(2\ell -n +1)}{2\ell}\qquad\qquad 2\le n< \ell+1.  
\end{eqnarray}
Hence the mass spectrum is the same as for the scalar fluctuations,
except
now the tachyon {\it and} the massless mode are missing.

In the $\ell\to\infty$ limit, after replacing $-\ln\phi-1/2$ by
$V'(\phi)/\phi$, we see that the relative coefficient between the last
two
terms in \eref{finallag} is a factor of two.  Examining the gauge-fixed
string
field action in \cite{KS}, one finds the same relative coefficient
between the
terms.

\sectiono{Concluding Remarks}\label{s7}

In this paper we have considered toy models for open string
field theory with actions
\begin{eqnarray}\label{fullaction}
{\cal S}_{\ell+1}
&=&\int d^Dx\Biggl\{-\, \half\partial_\mu\phi\partial^\mu\phi
-V_{\ell+1}(\phi)
-\half\partial_\mu A_\nu\partial^\mu A^\nu
\nonumber\\
&&\qquad\qquad\qquad
-\half \frac{{V_{\ell+1}}'(\phi)}{\phi}A_\mu A^\mu 
+\left(\partial_\mu\partial_\nu\ln\phi\right) A^\mu A^\nu\Biggr\},
\end{eqnarray}  
where
\begin{equation}
V_{\ell+1}(\phi)=\frac{\ell}{4}\phi^2\left(1-\phi^{2/\ell}\right).
\end{equation}
In the limit that $\ell\to\infty$, the action in \eref{fullaction} 
simplifies to
\begin{eqnarray}\label{infaction}
{\cal S}_{\infty}&=&\int d^Dx\Biggl\{
-\,\half\partial_\mu\phi\partial^\mu\phi
+\fourth\phi^2\ln\phi^2
-\, \half\partial_\mu A_\nu\partial^\mu A^\nu
\nonumber\\
&&\qquad\qquad\qquad
+\fourth\left(\ln\phi^2+1\right)A_\mu A^\mu 
+\left(\partial_\mu\partial_\nu\ln\phi\right) A^\mu A^\nu\Biggr\}.
\end{eqnarray}

The $\ell=\infty$ model in \eref{infaction}  
appears
to be more stringy than the finite $\ell$ models or the
$\ell=3$ model studied in \cite{0008227}. First, the
continuum spectrum is absent. Second, upon tachyon condensation
all modes disappear from the spectrum. Third, mass levels
are equally spaced and the tachyon on the lump has the same
mass as the tachyon in the original field theory model.
Fourth, higher codimension lumps satisfy simple descent
relations.

On the other hand the $\ell=\infty$ model also has some
features that are less similar to those in open string
field theory. In some sense $\ell=3$ matches quite nicely
with the cubic nature of open string field theory, and with the level
zero tachyon action.  In addition, tachyon condensation
in $\ell=3$ involves most of the spectrum just as in
string field theory, 
while in $\ell=\infty$ only the tachyon mode condenses.
Finally, being nonpolynomial, and having a level zero
tachyon field, it is not completely obvious how to deal
with the level expansion in the $\ell=\infty$ model. In
this respect, the model seems quite similar to closed
string field theory \cite{9206084}.
Learning how to do level expansion in the $\ell=\infty$
theory may eventually help to understand the still mysterious fate of
the closed string tachyon.

\bigskip

\noindent {\bf Acknowledgments}:
We thank J. Goldstone and R. L. Jaffe
for many conversations and detailed discussions on the subject
presented in this paper. The research of J.A.M. is supported in part by
NSF grant number PHY-97-22072 and by DOE contract 
\#DE-FC02-94ER40818.  The research of B.Z. is supported in part
by DOE contract \#DE-FC02-94ER40818.

\end{document}